\documentstyle[twoside,fleqn,espcrc2,epsfig]{article}

\title{Comparing lattice Dirac operators with Random Matrix
Theory\thanks{Supported by Fonds zur F\"orderung der  Wissenschaft\-li\-ch\-en
Forschung in \"Osterreich,  Project P11502-PHY.}}      

\author{F. Farchioni,  I. Hip\thanks{Poster presented by I. Hip}  and C.~B. Lang
\vspace{6pt}\\ {Institut f\"ur Theoretische Physik, Universit\"at Graz, A-8010
Graz, Austria}}

\begin{document}
\pagestyle{empty}

\begin{abstract}
We study the eigenvalue spectrum of different lattice Dirac operators
(staggered, fixed point, overlap) and discuss their dependence  on the
topological sectors. Although the model is 2D (the Schwinger
model with massless fermions) our observations indicate possible problems in 4D
applications. In particular misidentification of the smallest eigenvalues due
to non-identification of the topological sector may hinder successful
comparison with Random Matrix Theory (RMT). 
\end{abstract}

\maketitle

\section{INTRODUCTION}

In recent work we have been studying various aspects of the lattice  Schwinger
model \cite{FaHiLaWo,FaHiLa98}. This model is a 2D U(1) gauge theory of photons
and one or more fermion species. Of particular interest is the situation of
massless fermions. In the quantized theory chiral symmetry is broken by the
anomaly. The one flavor-model should exhibit a bosonic massive mode. 

For the non-perturbative lattice formulation chirality  is a central issue. The
Wilson Dirac operator explicitly breaks chiral symmetry. 
The Ginsparg-Wilson condition \cite{GiWi82} defines a class of
lattice actions with  minimal violation of chirality. An explicit realization
is Neuberger's overlap Dirac operator \cite{Ne98}. In another approach one 
attempts to construct so-called quantum perfect actions, or fixed
point actions (classically perfect actions) \cite{HaNi94}, 
also obeying the Ginsparg-Wilson condition \cite{Ha}.

%HaLaNi?

In the Schwinger model framework we have been studying several of these
suggestions. In \cite{LaPa98} the (approximate) fixed point Dirac operator was
explicitly  constructed. It has a large number of terms but has been shown to
have excellent scaling properties for the boson bound state propagators.  This
is not the case for the Neuberger operator \cite{FaHiLa98}; there scaling is
not noticeably improved  over the Wilson operator. The
overlap operator has eigenvalues distributed exactly on a unit circle
in the complex plane; for the (approximate) fixed point operator our study 
shows small (with smaller
$\beta=1/g^2$ increasing) deviations from exact circularity. In both cases we
could identify chiral zero modes. Their occurrence was strongly correlated to
the geometric topological charge of the gauge configuration
$\nu_\textrm{\footnotesize geo} = \frac{1}{2\,\pi}\sum_x
\textrm{Im}\,\textrm{ln} \, U_{12}(x)$ (henceforth called $\nu$ for brevity)
with a rapidly improving agreement with the Atiyah-Singer Index Theorem
(interpreted on the lattice) towards the continuum limit.

Studying the spectra of the Dirac operators suggests comparison with Random
Matrix Theory (RMT)\cite{GuMuWe98}. There the spectrum is separated in a
fluctuation part and a smooth background. Exact zero modes are disregarded. The
fluctuation part, determined in terms of the so-called unfolded variable (with
average spectral spacing normalized to 1), is conjectured to follow predictions
lying in one of three universality classes. For chiral Dirac operators these
are denoted by chUE, chOE and chSE (chiral unitary, orthogonal or symplectic
ensemble, respectively) \cite{RMT}.  Various observables have been
studied in this theoretical context. Comparison of actual data should verify
the conjecture and allows one to separate the universal features from
non-universal ones.  In particular it should be possible to determine in this
way the chiral condensate. 

On one hand the limiting value of the density for small eigenvalues and large
volume,
\begin{equation}
-\pi\,\lim_{\lambda\to 0}\lim_{V\to\infty} \rho(\lambda)=
\langle\bar\psi \psi\rangle\;,
\end{equation}
provides such an estimate due to the Banks-Casher relation. This information is
contained in the smooth average (background) of the spectral distribution.
However, also the fluctuating part, in particular the distribution for the
smallest eigenvalue $P(\lambda_\textrm{\footnotesize min})$ contains this
observable: Its scaling properties with $V$ are  given by unique functions of a
scaling variable $z\equiv \lambda V \Sigma$, depending on the corresponding
universality class. Usually this is the most reliable approach to determine
$\Sigma$, which then serves as an estimate for the infinite volume value of the
condensate in the chiral limit. This method does not involve unfolding,
averaging or extrapolation.

Here we concentrate on our results for the staggered Dirac  operator. It is
anti-hermitian and (for $m=0$) its spectrum is located on the  imaginary axis, but
it has no exact zero modes. RMT predictions for the staggered action and the
trivial topological sector have been confirmed also in 4D lattice studies
\cite{LGTRMT}. Here we emphasize, however, the r\^ole of non-zero topological
charge.

\section{METHOD AND RESULTS}

\begin{figure}[t]
%\vspace{-8pt}
\begin{center}
\epsfig{file=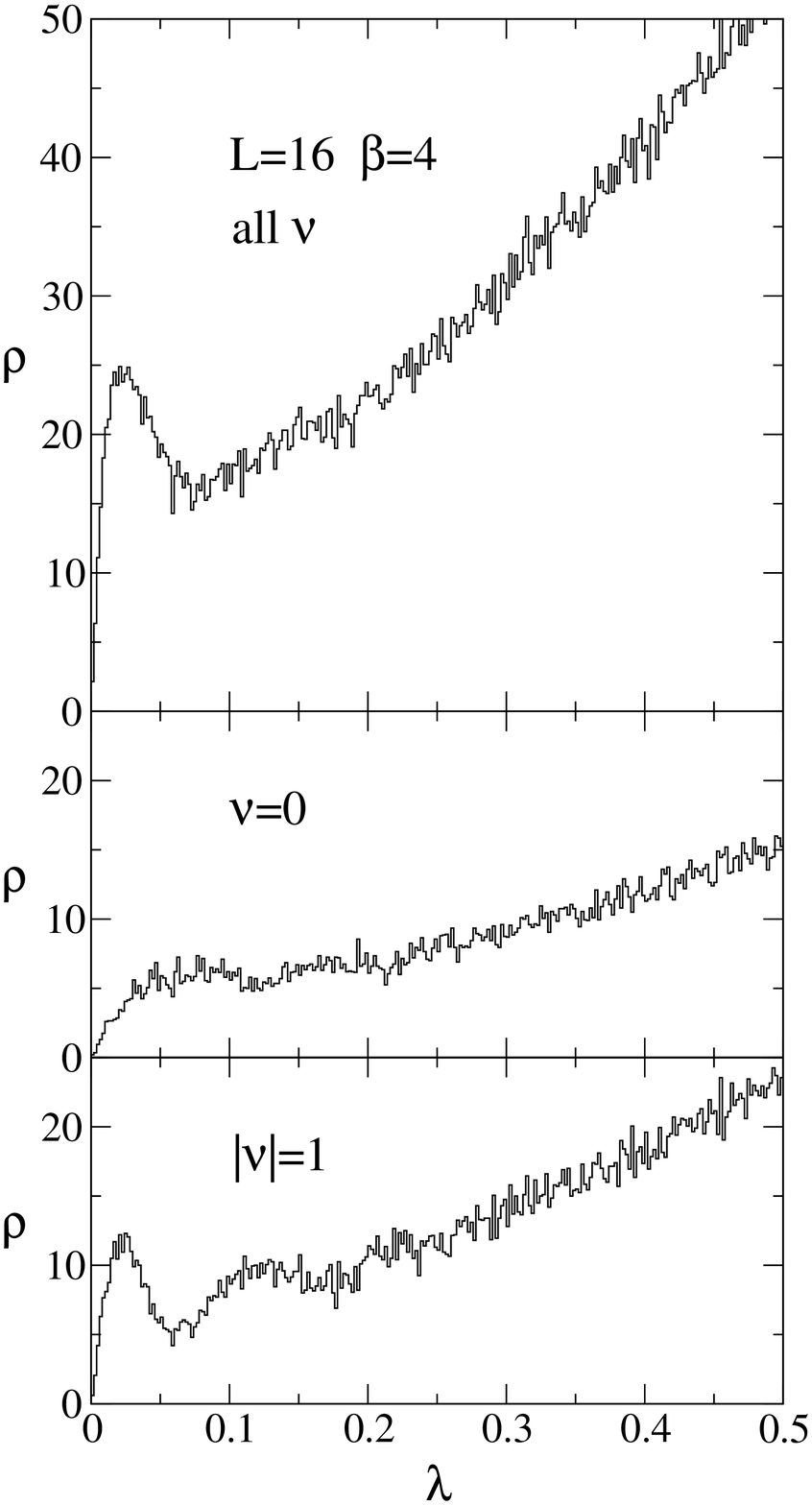,width=5cm}
\vspace{-28pt}
\end{center}
\caption{Distribution density of e.v.s of 
the staggered Dirac operator.
(out of all 10000 configurations 29\% have $\nu=0$  and 45\% have $|\nu|=1$.)}
\label{fig1}
\vspace{-12pt}
\end{figure}

In our study we construct sequences of (5000-10000) uncorrelated quenched gauge
configurations for several lattices sizes  ($16^2$, $24^2$, $32^2$) and  values
of $\beta$ (2, 4, 6). For these sets we then determine the various Dirac
operators and study their spectral distribution.  This way we can compare
directly the effect of identical sets of gauge configuration on the fermionic
action.  In \cite{FaHiLaWo} we discuss our results for the Neuberger- and the
fixed point operator. Since these spectra lie on or close to a circle in the
complex plane, one has to project them to the (tangential) imaginary axis.  We
find that they exhibit the universal properties of the (expected) chUE-class,
unless the physical lattice volume is too small. 

\begin{figure*}[th]
\vspace{-8pt}
\begin{center}
\epsfig{file=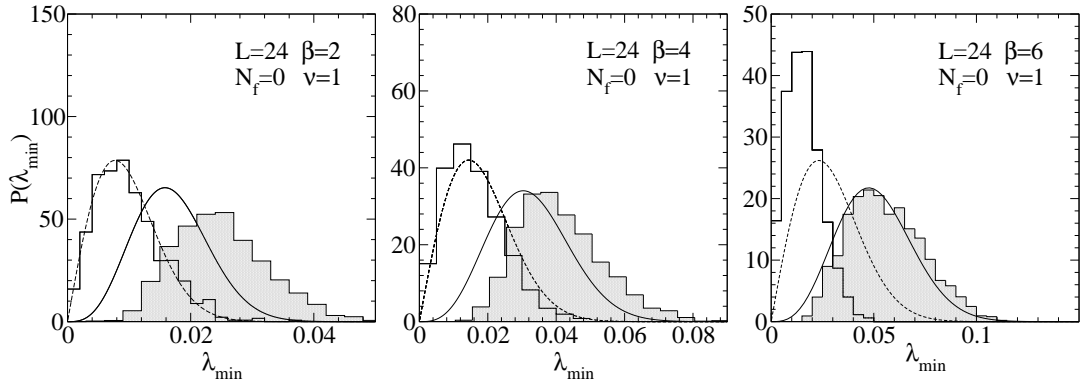,height=5cm}
\vspace{-28pt}
\end{center}
\caption{Distribution for the smallest and the 2nd smallest (shaded histogram) 
eigenvalues {\em in the $\nu=1$ sector}, quenched ($N_f=0$). The curves give
the chUE predictions for the smallest e.v. in the $\nu=0$ sector (dashed line)
in the $\nu=1$ sector (full line). }
\label{fig2}
\vspace{-12pt}
\end{figure*}

In Fig. \ref{fig1} we demonstrate the relevance of topological modes.  The e.v.
distribution density is first shown without distinguishing between different
$\nu$ and we notice a pronounced peak at small eigenvalues. Splitting the
contributions according to $|\nu|=0$ and 1 we observe, that the peak is due to 
the non-trivial sectors $\nu\neq 0$. The trivial sector has a behavior typical
for the shapes predicted from chRMT. For larger $\beta$ and $V$ the peak
becomes more pronounced, justifying the hypothesis that it represents the
``would-be'' zero modes.

Since RMT discusses the distribution {\em excluding exact zero-modes} we expect
problems whenever one is in a situation without possibility to separate
topological sectors (upper-most figure in Fig. \ref{fig1}), if one then tries to
represent the distribution for the smallest observed eigenvalue  by chRMT
functions. This is demonstrated in Fig. \ref{fig2} where we plot the histograms
for the smallest and the 2nd smallest (shaded histogram)  eigenvalues {\em in
the $|\nu|=1$ sector}. For small $\beta$,  strong coupling, the histogram for
the smallest e.v. behaves like the  $\nu=0$ sector prediction. For large
$\beta$ the 2nd smallest e.v.  follows a distribution expected for the {\em
smallest} e.v. in the  $|\nu|=1$ sector.

\begin{figure}[h]
%\vspace{-12pt}
\begin{center}
\epsfig{file=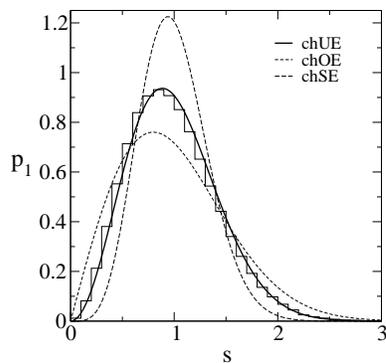,width=5cm}
\vspace{-28pt}
\end{center}
\caption{Level spacing distribution (for the unfolded variable $s$) 
for $L=16$, $\beta=4$,
compared to chRMT predictions for chUE (full), chOE (dashed) and chSE 
(dotted).}
\label{fig3}
\vspace{-12pt}
\end{figure}

The level spacing distribution (determined in the unfolded variable) clearly
has chUE (Wigner surmise) shape (Fig. \ref{fig3}) for all sizes and $\beta$.

Having all eigenvalues we can of course calculate the fermion determinant for
every gauge configuration and include dynamic fermions by explicit
multiplication. These ``unquenched'' results will be presented elsewhere.


\begin{thebibliography}{9}

\bibitem{FaHiLaWo}
F. Farchioni, I. Hip, C.~B. Lang and M. Wohlgenannt,
\newblock Nucl. Phys. B 549 (1999) 364 and references therein.

\bibitem{FaHiLa98}
F. Farchioni, I. Hip, and C.~B. Lang,
\newblock Phys. Lett. B 443 (1998) 214.

\bibitem{GiWi82}
P.~H. Ginsparg and K.~G. Wilson,
\newblock Phys. Rev. D 25 (1982) 2649.

\bibitem{Ne98}
H. Neuberger,
\newblock Phys. Lett. B 417 (1998) 141; ibid. 427 (1998) 353.

\bibitem{HaNi94}
P. Hasenfratz and F. Niedermayer,
\newblock Nucl. Phys. B  414 (1994) 785.

\bibitem{Ha}
P. Hasenfratz,
\newblock Nucl. Phys. B (Proc. Suppl.) 63A-C (1998) 53;
Nucl. Phys. B 525 (1998) 401.

\bibitem{LaPa98}
C.~B. Lang and T.~K. Pany,
\newblock Nucl. Phys. B 513 (1998) 645.

\bibitem{GuMuWe98}
T. Guhr, A. M{\"u}ller-Groeling, and H.~A. Weidenm{\"u}ller,
\newblock Phys. Rep. 299 (1998) 189.

\bibitem{RMT}
H. Leutwyler and A. Smilga,
\newblock Phys. Rev. D 46 (1992) 5607.
E.~V. Shuryak and J.~J.~M. Verbaarschot,
\newblock Nucl. Phys. A 560 (1993) 306;
J.~J.~M. Verbaarschot,
\newblock Phys. Rev. Lett. 72 (1994) 2531;
P.~H. Damgaard, \newblock Phys. Lett. B 424 (1998) 322;
G. Akemann and P.~H. Damgaard, \newblock Nucl. Phys. B 528 (1998) 411;
J.~C. Osborn et al., Nucl. Phys. B 540 (1999) 317;
P.~H. Damgaard et. al.
Nucl.Phys. B 547 (1999) 305.

\bibitem{LGTRMT}
M.~E. Berbenni-Bitsch et~al.,
\newblock Phys. Rev. Lett. 80 (1998) 1146;
J-Z. Ma et. al.,
\newblock Eur. Phys. J. A2 (1998) 87;
M.~E. Berbenni-Bitsch et~al.,
\newblock Phys. Rev. D 58 (1998) 071502;
P.~H. Damgaard, U.~M. Heller, and A. Krasnitz,
\newblock Phys. Lett. B 445 (1999) 366;
M. G\"ockeler et. al.,
\newblock Phys. Rev. D 59 (1999) 094503.

\end{thebibliography}
\end{document}